\documentclass[12pt]{article}
\usepackage[utf8]{inputenc}
\usepackage{hyperref}                       
\usepackage{setspace}
\usepackage{palatino}                       
\usepackage{graphicx}
\usepackage{xcolor}                         
\usepackage{float}  
\usepackage{multirow}   
\usepackage{lscape} 
\usepackage[total={6in, 9.5in}]{geometry}   
\usepackage{microtype}                      
\microtypesetup{
  protrusion=true,
  expansion=true
}                                           
\usepackage{amsmath}                        
\usepackage{amssymb}                        
\usepackage{amsthm}                         
\newtheorem{definition}{Definition}         
\newtheorem{theorem}{Theorem}               
\newtheorem{corollary}{Corollary}           
\usepackage[flushleft]{threeparttable}      
\usepackage{subcaption}                     
\usepackage{tocbibind}
\usepackage[toc,page]{appendix}             
\fontfamily{ppl}\selectfont 
\usepackage[american]{babel}                
\usepackage{setspace}

\usepackage{natbib} 
\fontfamily{ppl}\selectfont 
\usepackage[american]{babel}
\usepackage{setspace}

\usepackage{titling} 
\title{Theoretical Steps to Optimize Transportation in the Cubic Networks and the Congestion Paradox
\thanks{Presented on Econ 644 class in the Econ PhD Program in Rutgers University on Dec. 12, 2023 }
\thanks{Special thanks to Professor Barry Sopher for his insightful comments and suggestions that greatly improved this paper. Thanks to Lucas H. Lee for his flourish suggestion that improved expressions in this paper. And, thanks to my partner Arin Jeong for her invaluable contribution and support providing encouragement throughout the research process.}
}

\author{Joonkyung Yoo\\
    \href{mailto:yhsgenie@gmail.com}{\texttt{yhsgenie@gmail.com}}
    \quad
    \href{mailto:jy768@rutgers.edu}{\texttt{jy768@rutgers.edu}}
    }
    
\date{\today}

\begin{document}


{\setstretch{1.0}
\maketitle

\begin{abstract}
 Given a player is guaranteed the same payoff for each delivery path in a single-cube delivery network, the player's best response is to randomly divide all goods and deliver them to all other nodes, and the best response satisfies the Kuhn-Tucker condition. The state of the delivery network is randomly complete. If congestion costs are introduced to the player's maximization problem in a multi-cubic delivery network, the congestion paradox arises where all coordinates become congested as long as the previous assumptions about payoffs are maintained.
  



\end{abstract}

}


\section{Introduction}

What should be the traffic signals for airborne vehicles in the near future? What should be the principles of a universal airborne transportation system that allows both of unmanned aerial vehicles and manned aerial vehicles to travel simultaneously? The introduction of drones marked the beginning of a generalized use of airspace previously reserved for airplanes, helicopters, or other high-speed fighter jets for specific military purposes, as well as for various projectiles and rockets. The use of drones for large-scale sporting events such as the Olympics is no surprise, and a company in South Korea has launched drone services to deliver emergency packages to the Yeo-su Islands (\cite{osullivan-dale_skyports_2023}). And Amazon in the United States has introduced and is implementing drone deliveries in some areas (\cite{bbc_news_how_2023}). Meanwhile, manned aerial vehicles, which are slightly larger human-manned versions of drones, are being developed in South Korea and the United States. Perhaps in the not-too-distant future, one will encounter the paths of aerial vehicles more often than observing the complex paths of satellites in the sky.

The Line in Saudi Arabia, a project that is even bigger than a skyscraper, is an immeasurable structure that requires not only the movement of mobility, but the placement of elevators and escalators in the interior of a enormous human-made structure. Designing the internal circulation of the building paths including diagonal movement will dramatically reduce the cost of transportation. However, considering the intersection of the diagonal paths, what are the efficient principles of construction design for escalators and elevators?

This research started with the question of traffic rules for such aerial vehicles and the optimal transportation paths in three-dimensional spaces such as the internal routes of mega-structures. First of all, one needs an airborne transportation system to minimize the transaction cost at intersections such as traffic lights and stop signs for ground vehicles.

In the case of mega-structures, the design of multiple diagonal escalators intersecting at a single point or the design of an elevator center already exists in skyscrapers and large square centers. However, some travel mezzanines are complex, twisted, and sometimes blocked by renovations or other reasons, increasing the cost of travel.

This study aims to discover general rules for optimal travel routes in three-dimensional space using network theory.  In particular, my goal is to build some first steps in understanding what the efficient transportation rules in a three-dimensional space must be when there are many objects that want to get from one point to another. Ultimately, it is to answer the question of what causes inefficiency in traveling in three-dimensional space.

Therefore, this research focuses on two milestones on travel paths in a three-dimensional space. First, what model should be used to optimize the benefits and costs of goods originating from each node? Second, if there are congested intersections, how can it consider them for the utility maximization of each player to find the optimal route? To this end, this thesis seeks to devise and optimize the payoff function of individual nodes whose utility is maximized upon the completion of the transport. At the same time, we graphed all the possible cases where there would be an intersection in the path of the two combined cubes.

Network theory and optimal transport are the literature of economics, mathematics, computer science, and transportation engineering that provides effective theoretical and empirical support to answer this question. Including the seminal book such as \cite{rachev_mass_1998} in transportation engineering, the classics of optimal transport by the Field's Medal mathematician, C\'edric Villani (\cite{villani_optimal_2009}) and other literatures on optimal transport, \cite{villani_topics_2003}, \cite{ekeland_optimal_2010}, \cite{ekeland_comonotonic_2012}, \cite{santambrogio_optimal_2015}, and \cite{carlier_vector_2016}, have flourished in economics with the application of optimal transport in economics such as \cite{ekeland_optimal_2010} and \cite{galichon_optimal_2016}. Since then, numerous papers have been published on optimizing the movement of vehicles such as taxis and bus ambulances in two-dimensional spaces such as the Manhattan road network.

However, compared to these theoretical and empirical evidences to create rules for ground transportation, the large-scale transportation rules for the airborne vehicles must be far more strict and restrained. What if all the traffic lights and stop signs are hung on an aerial roadway where Urban Air Mobility (UAM) and Advanced Air Mobility (AAM) are traveling simultaneously, where must they be placed? It is silly to think of traffic rules that assume a large number of traffic lights and stop signs that hang a mile above the ground. It is not much different when you have an overabundance of traffic islands and terminals high up in mega-structures, blocking sunlight and visibility.

Therefore, in this paper, we started from the beginning and modeled the player's optimization problem of the delivery path in a three-dimensional cube space by applying network theory. We defined congestion and considered the possibility of a three-dimensional path that avoids the congestion.

The remainder of the paper is organized as follows. In Chapter 2, we evolved the delivery models from linear networks to cubic networks and analyze each delivery model. In Chapter 3, we defined the congestion that arises in the multi-cube model and visualized the picture in which this congestion occurs. We also proposed ways to solve the congestion and provided all two-cube examples including congestion. Chapter 4 further discussed the problems that arise when costs are defined as a distance and divergence function, considered a social planner model that can supply roads, and concluded.

\section{Model}

We define a delivery model and discuss utility maximization to progress from a linear model to a three-dimensional model to simplify analysis.

\subsection{The linear network and the plane Network}

Let $i$ $\in N=\{1,2\}$ be a player. In the model player $i$ has the benefit $b_{ij}$ of delivering goods to the other nodes $j$. The player $i$ cannot have any benefit if she does not deliver the goods to the other player. Thus, the benefit of storing goods $b_{ii}$ is zero. In this model, we assume that the storage of goods is positively cost $c_{ii} > 0$. Each player $i$ chooses a level of goods $x_{ij} \in  X_i = [0,1]$ to deliver to the player $j$ such that $\sum_{j}x_{ij} = 1$. The set of links $g_{ij} \in G_i = \{0,1\}$. The path $v_{ij}$ exists if the link $g_{ij} = 1$. Given the link $g_{ij}$ to get to the other node, each player $i$ also chooses a path $v_{ij}$ to deliver to the player $j$, which is $v_{ij} \in V_i = \{0,1\}$, where $i \neq j$. Thus, we say that the player $i$ uses a delivery path $v_{ij}$ if $v_{ij} = 1$. Let define $v_{ii}$\footnote
    {\, The $v_{ii}$ is an imaginary vector of the player $i$'s storage decision, which is either one or zero. It is one if the player $i$ chooses a positive amount of goods to store in his storage, the original node $i$, then the storage cost term in the profit function, $\Pi_i$, must be reflected, which is multiplied by one, $v_{ii}$. The positive amount of goods must be dumped if $v_{ii} = 0$ but $x_{ii}>0$. This case is ignored in this network theory because the goods must not be dumped out. And then $v_{ii}$ is zero, reflecting the removal of the storage cost term if there are no goods to store at the original node.
} and $v_{ij}$ as:
\begin{equation}
    v_{ii} =
        \begin{cases} 
            1 & \text{if } x_{ii} > 0\\
            0 & \text{if } x_{ii} = 0
        \end{cases},
        \quad 
    v_{ij} =
        \begin{cases} 
            1 & \text{if } x_{ij} > 0\\
            0 & \text{if } x_{ij} = 0
        \end{cases} 
\end{equation}
.\footnote{\, By the definition, there exist four different vectors $\textbf{v}_i$ which correspond to vector $\textbf{x}_i$.

If $\textbf{v}_i = \begin{bmatrix}v_{ii} \\ v_{ij} \end{bmatrix}$ is $\begin{bmatrix}0 \\ 0 \end{bmatrix}$, then $\textbf{x}_i = \begin{bmatrix}x_{ii}=0 \\ x_{ij}=0\end{bmatrix}$. Similarly, $\textbf{v}_i = \begin{bmatrix}1 \\ 1 \end{bmatrix} \Longleftrightarrow \textbf{x}_i = \begin{bmatrix}x_{ii}>0 \\ x_{ij}>0\end{bmatrix}$, $\textbf{v}_i = \begin{bmatrix}0 \\ 1 \end{bmatrix} \Longleftrightarrow \textbf{x}_i = \begin{bmatrix}x_{ii}=0 \\ x_{ij}>0\end{bmatrix}$, and $\textbf{v}_i = \begin{bmatrix}1 \\ 0 \end{bmatrix} \Longleftrightarrow \textbf{x}_i = \begin{bmatrix}x_{ii}>0 \\ x_{ij}=0\end{bmatrix}$} Then, the values in the path vector indicate whether the goods to stock or deliver are positive or not. Furthermore, if a link does not exist, the player cannot choose the path, which is $v_{ij} \leq g_{ij}$. The linear network is shown in Figure \ref{fig:1} below.
\begin{figure}[H]
    \centering
        \includegraphics{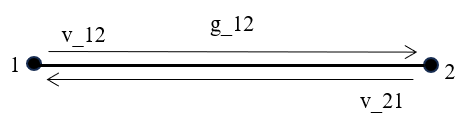}
    \caption{The linear network}
    \label{fig:1}
\end{figure}

In this linear network, we define that the utility function $\textbf{U}_1(x_{12}, v_{12}): [0,1] \times \{0,1\} \to \mathbb{R}$ is a combination of the benefit $b_{ij}$, the delivery path $v_{ij}$, and the amount of delivery goods $x_{ij}$: $\textbf{U}_i(x_{ij}, v_{ij}) = b_{ij}v_{ij}x_{ij}$ with $i \neq j$. The benefit is greater than the cost in the delivery process, and the cost of storage is always positive. The set of strategies of the player $i$ is denoted by $S_i = X_i \times V_i$. Define $S = S_1 \times S_2$ as the set of strategies of all players. A strategy profile $\textbf{s}= (\textbf{x}, \textbf{v}) \in S$ acquired by each player specifies the vector of goods delivered and the vector of the delivery path, $\textbf{x} = (x_1,x_2)$ and $\textbf{v} = (v_1,v_2)$, respectively. Thus, the player $1$ maximizes his payoff:
\begin{equation}
    \max_{S_{1}(\textbf{x}_{1},\textbf{v}_{1})}\Pi_1 = \textbf{U}_1({x}_{12}, {v}_{12})- \begin{bmatrix}{v}_{11} \\ {v}_{12}\end{bmatrix}' \begin{bmatrix}c_{11} & 0 \\ 0 & c_{12}\end{bmatrix} \begin{bmatrix}{x}_{11} \\ {x}_{12}\end{bmatrix}
\end{equation}
such that
\begin{equation}
\begin{aligned}
    {c}_{ii} &> 0, \qquad \forall \; i \in N \\
    b_{ij} > c_{ij} &> 0, \qquad i \neq j\\
    \text{and} \;\sum_{j \in N}x_{ij} &= 1.
\end{aligned}
\end{equation}
Player 2 maximizes his payoff analogously. Then the equilibrium\footnote{\, Refer to Appendix \ref{apdx:A} about the Kuhn-Tucker conditions of all of strategies of the player $i$.} 
\begin{equation}
    \textbf{U}_i(\cdot, \cdot) = (b_{ij} - c_{ij})v_{ij}x_{ij}, \;\text{under} \; S_i = \begin{bmatrix}1 \\ 0 \end{bmatrix} \times \begin{bmatrix}1 \\ 0 \end{bmatrix} \; \text{for} \; i. 
\end{equation}
Hence, the player in the linear network always chooses to deliver all amount of goods to the opponent node to maximize his utility in this model.

Similarly, the players in the plane network also behave to maximize his payoff. Let $i$ $\in N=\{1,2,3,4\}$ be a player. Without loss of generality, the player $i$ also chooses a set of paths to deliver to others, represented as a vector $\textbf{v}_i = [v_{ij}, v_{ij}] \in V_i = \{0,1\}^4$, where $i \neq j$, and a set of goods to deliver to the other three nodes, a vector $\textbf{x}_i = [x_{ij} \in [0,1], x_{ij} \in [0,1]] \in X_i = [0,1]^4$, where $i \neq j$. Then the difference from the linear model is the domain of the player's utility function. We define the utility function in the plane network $\textbf{U}_i(\cdot,\cdot): X_i \setminus \{x_{ii}\} \times V_i \setminus \{v_{ii}\} \to \mathbb{R} = \sum_{j \neq i}{b_{ij}v_{ij}x_{ij}}$. To simplify the equilibrium in this network, we assume that there are no differences in benefit and cost from the delivery using through between the diagonal path or the normal path. That is, all $c_{ij} \; (j \neq i)$ cost equivalently, and all benefits $b_{ij} \; (j \neq i)$ are the same. Thus, the equilibrium $S = S_1 \times S_2 \times S_3 \times S_4$ in this complete plane network is achieved when all players maximize their payoff. The plane network is shown in Figure \ref{fig:2} below.
\begin{figure}[H]
    \centering
        \includegraphics{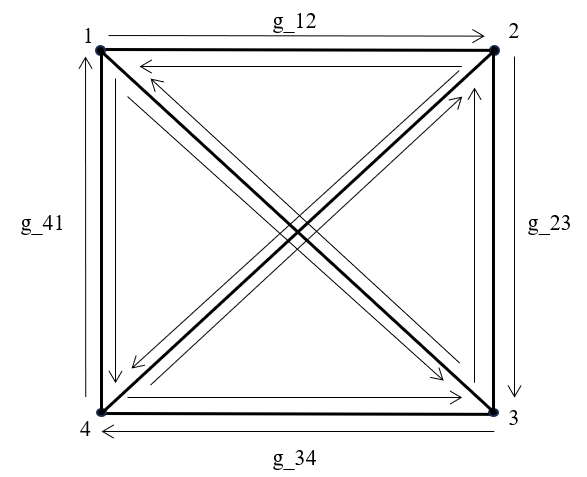}
    \caption{The plane network}
    \label{fig:2}
\end{figure}

Player $i$ maximizes his payoff:
\begin{equation}
\max_{\textbf{x}_{i \neq j}, \textbf{v}_{i \neq j}}\Pi_i = \textbf{U}_i(\textbf{x}_{i \neq j}, \textbf{v}_{i \neq j})- \textbf{v}_{i} ' \begin{bmatrix}c_{i1} & 0 & 0 & 0\\ 0 & c_{i2} & 0 & 0 \\ 0 & 0 & c_{i3} & 0 \\ 0 & 0 & 0 & c_{i4} \end{bmatrix} \textbf{x}_i \text{ for some } i, j \in N
\end{equation}
such that,
\begin{equation}
\begin{aligned}
    {c}_{ii} &> 0, \qquad \forall \; i \in N \\
    b_{ij} > c_{ij} &> 0, \qquad i \neq j\\
    \text{and} \;\sum_{j \in N}x_{ij} &= 1.
\end{aligned}
\end{equation}
The other players maximize their payoffs analogously. Because the structures of the costs and benefits of the delivery are the same as the linear network, the best response of the player $i$ is $S_i = \textbf{x}_i \times \textbf{v}_i$ such that some $x_{ij}>0$ under $\sum_{j \in N}x_{ij} = 1$ and $x_{ii} = 0$, and the path choices $v_{ij} = 1$ and $v_{ii} = 0$ for $j \neq i$. That is, the player in the plane network delivers all the goods to the other nodes. In this delivery decision, the arrival nodes and the amount of the arrival goods are random. This plane network is complete the reason why we assume the same benefit and the same cost for delivery goods. If the assumption is breakdown, we will delve into the case in the discussion chapter.

\begin{theorem}
    In a well-defined linear and plane delivery network under linear utility and cost functions, the plane delivery network is {\large\textbf{randomly complete}} if the cost of holding the goods is positively the same, and the payoff of the delivery is positively the same for all players.
    \label{Thm:1}
\end{theorem}
\begin{corollary}
    All players in the randomly complete delivery network have same best response if and only if the same Kuhn-Tucker condition in their payoff maximization problems.
    \label{crl:1}
\end{corollary}

\subsection{The cubic Network}

When defining each node in a three-dimensional model, unlike the one-dimensional and two-dimensional models, we need to use three different notations for numbering so that we can associate information about the location of the node with its name. Although the node notation follows a known numbering up to the above delivery model, the three-dimensional model will use a new node notation. Let $i$ $\in N=\{{l,m,n}\}$ for $l, m, n \in \{0,1\}$ be a player. In other words, the set of players in a three-dimensional network is defined by their coordinates in the node notation. For example, a player is noted as $i_{0,0,0}$ and another player is noted as $i_{0,0,1}$, which means that both players are distanced by one on the third axis in this network. We also note that the link for two nodes $i_{l,m,n}$ and $i_{l',m',n'}$ is $g_{ll',mm',nn'}$ for $l \neq l', m \neq m', \text{ and } n \neq n'$, and the delivery path from $i_{l,m,n}$ to $i_{l',m',n'}$ is noted as $v_{ll',mm',nn'}$. On the contrary, the path of delivery from $i_{l',m',n'}$ to $i_{l,m,n}$ is noted as $v_{l'l,m'm,n'n}$. Similarly, the goods to be delivered from $i_{l,m,n}$ to $i_{l',m',n'}$ is noted as $x_{ll',mm',nn'}$, and from $i_{l',m',n'}$ to $i_{l,m,n}$, as $x_{l'l,m'm,n'n}$ for $l \neq l', m \neq m', \text{ and } n \neq n'$. 

Without loss of generality, the player $i_{l,m,n}$ chooses a set of paths to deliver to others, represented as a vector $\textbf{v}_i = [v_{ll,mm,nn}, v_{ll',mm',nn'}] \in V_i = \{0,1\}^8$ in this cube network, where $l \neq l', m \neq m', \text{ and } n \neq n'$, and a set of goods to deliver to the other nodes, a vector $\textbf{x}_i = [x_{ll,mm,nn} \in [0,1], x_{ll',mm',nn'} \in [0,1]] \in X_i = [0,1]^8$. Then the difference from the plane model is the domain of the player's utility function. We define the utility function in the cubic network $\textbf{U}_i(\cdot,\cdot): X_i \setminus \{x_{ll,mm,nn}\} \times V_i \setminus \{v_{ll,mm,nn}\} \to \mathbb{R} = \sum_{m \neq l}\sum_{n \neq l}{b_{ll',mm',nn'}v_{ll',mm',nn'}x_{ll',mm',nn'}}$. To simplify the equilibrium in this network, we assume that there are no differences in benefit and cost from the delivery using through between the diagonal path or the normal path. That is, all $c_{ll',mm',nn'} \; (l \neq l', m \neq m', \text{ and } n \neq n')$ cost equivalently, and all benefits $b_{ll',mm',nn'} \; (l \neq l', m \neq m', \text{ and } n \neq n')$ are the same. Define the strategy function of the player $i_{l,m,n}$ as $S_{l,m,n}$. Then, equilibrium $S = S_{0,0,0} \times \ldots \times S_{1,1,1}$ in this complete cubic network is achieved when all players maximize their payoff. The cubic network is drawn as in Figure \ref{fig:3} below. 

\begin{figure}[H]
    \centering
        \includegraphics{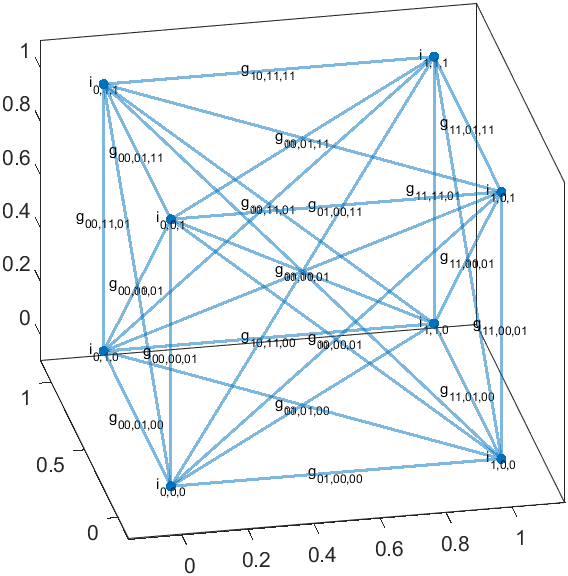}
    \caption{The cubic network}
    \label{fig:3}
\end{figure}

Player $i$ maximizes his payoff:
\begin{equation}
\begin{aligned}
    \max_{\textbf{x}_{l,m,n}, \textbf{v}_{l,m,n}}\Pi_i &= \textbf{U}_{l,m,n}(\cdot, \cdot)- \textbf{v}_{l,m,n}' \, \mathbb{C} \, \textbf{x}_{l,m,n},\\
    \text{ and } \mathbb{C} &= c_{l,m,n}' \, \mathbb{I} \text{ for some } l, m, n \in N
\end{aligned}
\end{equation}
such that,
\begin{equation}
\begin{aligned}
    c_{ll,mm,nn} &> 0, \qquad \forall \; l, m, n \in N \\
    b_{ll',mm',nn'} > c_{ll',mm',nn'} &> 0, \qquad l \neq l', m \neq m', \text{ and } n \neq n'\\
    \text{and} \;\sum_{l}\sum_{m}\sum_{n}x_{ll',mm',nn'} &= 1.
\end{aligned}
\end{equation}
The other players maximize their payoffs analogously. All players have the same payoff maximization problem in this cubic network, which means that the K-T conditions of all the maximization problems are the same. By Corollary \ref{crl:1}, this cubic network is \textit{randomly complete}.

\section{Congestion in the multiple cubic Network}

\subsection{Congestion}

We have considered the payoff maximization problem for a single cube delivery network. We consider the same payoff maximization problem for a delivery network with two cubes in this chapter; however, there is a complicated issue to think about when maximizing in a multi-cube delivery network: congestion.

\begin{definition}
    {\large\textbf{Congestion}} is a state of increasing cost in the space of the delivery network where two different delivery paths pass temporarily or continuously through the same single or multiple coordinates.
    \label{def:1}
\end{definition}

We can observe from Figure \ref{fig:3} that intersections exist in three-dimensional traffic, just as they do on the ground. On one hand, congestion inside a single cube structure is solved in the real world situation as follows. On the ground, intersections are known to minimize travel transaction costs by using traffic lights and STOP signs. A classic example of transaction cost minimization in three-dimensional space is an airport control tower. In a large building, such as a department store, a fixed center of elevators and escalators could play this role. Despite the fact that we can already observe a large number of intersections in the previous single-cube network model, we have not taken into account these transaction costs that may arise because of congestion. On the other hand, when the delivery network is organized into two cubes, another congestion problem arises: the delivery paths of different pairs of nodes overlap. An airport control tower solves this problem by assigning an order to the planes that need to take the runway. Buildings solve this problem by having multiple lines of elevators and escalators. This phenomenon of overlapping paths is known as a bottleneck.

To solve these congestion issues, the first general rule is to give each delivery an occupancy permit. In other words, a single delivery path can exclusively occupy only one intersection or one link. While congestion in two dimensions can be solved through transaction costs such as traffic lights or STOP signs, congestion in three dimensions can be solved by stopping in mid-air, which is either impossible or energy-consuming, so "occupancy permits" are the most viable alternative. Therefore, we consider the applicability of these occupancy permits in a payoff maximization problem. That is, we delve into if congestion costs as additional cost term in the payoff function. Define the two types of congestion: the intersection and the bottleneck.

\begin{definition}
    {\large\textbf{Point congestion}} is the congestion at a specific single coordinate, not a node, where two or more links intersect at a point and the delivery paths cross each other.
    \label{def:2}
\end{definition}

Point congestion is mainly caused by plane diagonals or space diagonals, so it has occurred since the two-dimensional delivery network. In the plane network as Figure \ref{fig:2}, we can already observe the intersection within the plane.

\begin{definition}
    {\large\textbf{Line congestion}} is the lasting congestion along a single unit link, where one link overlaps another link, so two different delivery paths are sufficiently shared on the single unit link.
    \label{def:3}
\end{definition}

Line congestion mostly occurs at the edges where two cubes meet, but it can also occur on the diagonal in more than three cubes.

In practice, the occupancy permit could be the alternative for congestion costs; however, the theoretical problem of assigning one path to one congestion is not as simple as it sounds. To do so, we need to know which type congestion appears and how many the number of each congestion increases along with the number of cubes. Discovery of the increasing sequence of the congestion number is as difficult as finding a particular sequence in geometry or number theory, because the number of point congestions and line congestions proliferates with each additional cube notwithstanding the cubic network is geometrically simple.

There is another challenge: the difficulty of the optimization including congestion costs. According to the Definition \ref{def:1}, it costs at a specific coordinate, which means each player must optimize his payoff in terms of the congestion cost all over coordinates. Since every link connected to a player is exposed to line congestion and point congestion, if the congestion cost $\varepsilon >0$, the cost will grow to infinity. Imagine a node in a cube surrounded by 26 other cubes. The player at the node is exposed congestion on all links including diagonals, which makes the optimization difficult. We shall see this optimization is impossible for the later part.

Scaling in cubic networks can be categorized in three ways: plane, edge, and node sharing, which increase the total number of nodes in this order. We explore how congestions arises in each type of network in order.

\subsection{The cubic network sharing plane, edge, and Node}

\begin{figure}[H]
   \centering
       \includegraphics[scale=.7]{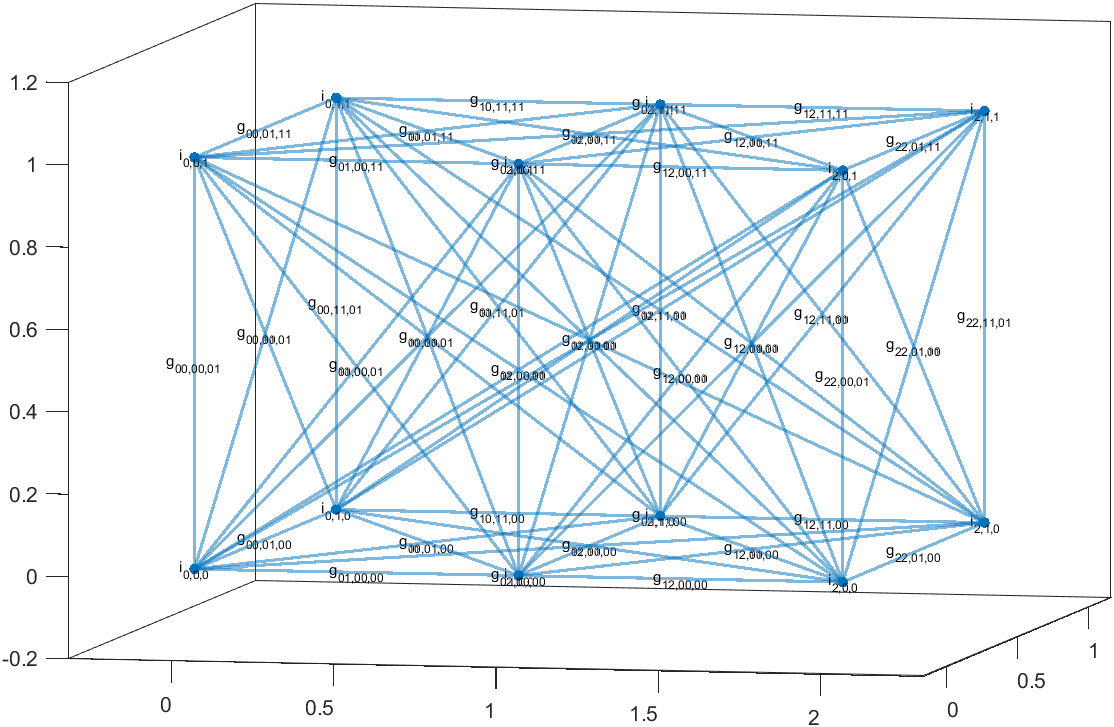}
   \caption{The complete plane-sharing cubic Network}
   \label{fig:4}
\end{figure}

We construct a plane-sharing cubic network using the nodes and links we defined in the single cube delivery network. In this cube, there are a total of 66 links: 20 unit links, 22 planar diagonal links, 8 spatial diagonal links, 8 long planar diagonal links, 4 long spatial diagonal links, and 4 long edge links. This shared cubic network is drawn in Figure \ref{fig:4}.

\begin{figure}[H]
   \centering
       \includegraphics[scale=.7]{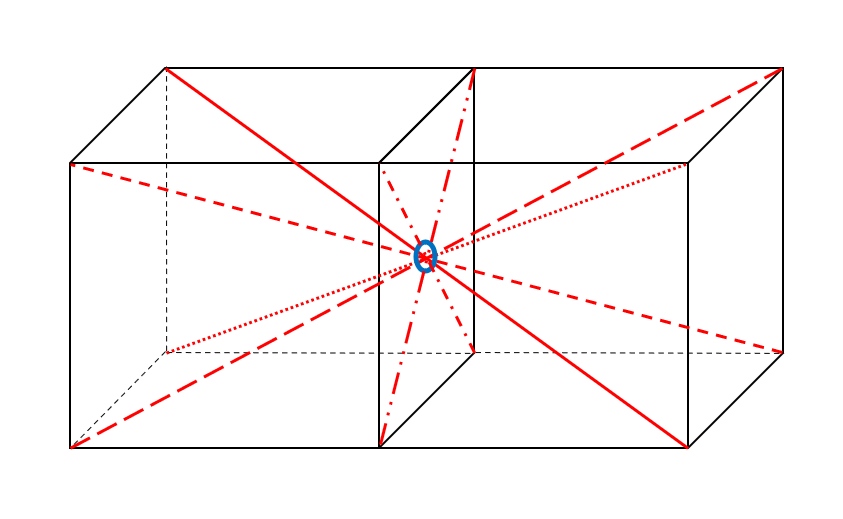}
   \caption{The unique point congestion in the plane-sharing cubic Network}
   \label{fig:5}
\end{figure}

\begin{figure}[H]
   \centering
       \includegraphics[scale=.7]{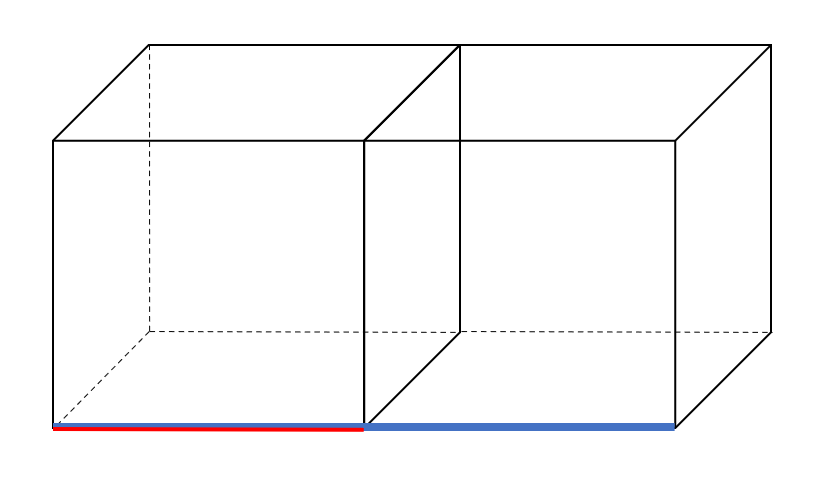}
   \caption{A line congestion Example}
   \label{fig:6}
\end{figure}

In this plane-sharing two-cubic network, a unique \textit{redundunt} point congestion appears, where the new point congestion in two-cubic space overlaps the old point congestion in the shared plane. The location of this point congestion is shown in Figure \ref{fig:5}. On the other hand, this plane-sharing two-cubic network allows us to explore the possibility of line congestion to a very large extent: if a sufficient number of cubes are connected to each other, we can conjecture that almost all unit links will be line congested. Figure \ref{fig:6} shows one of the eight possible line congestions in this plane-sharing two-cubic network.

\begin{figure}[H]
   \centering
       \includegraphics[scale=.7]{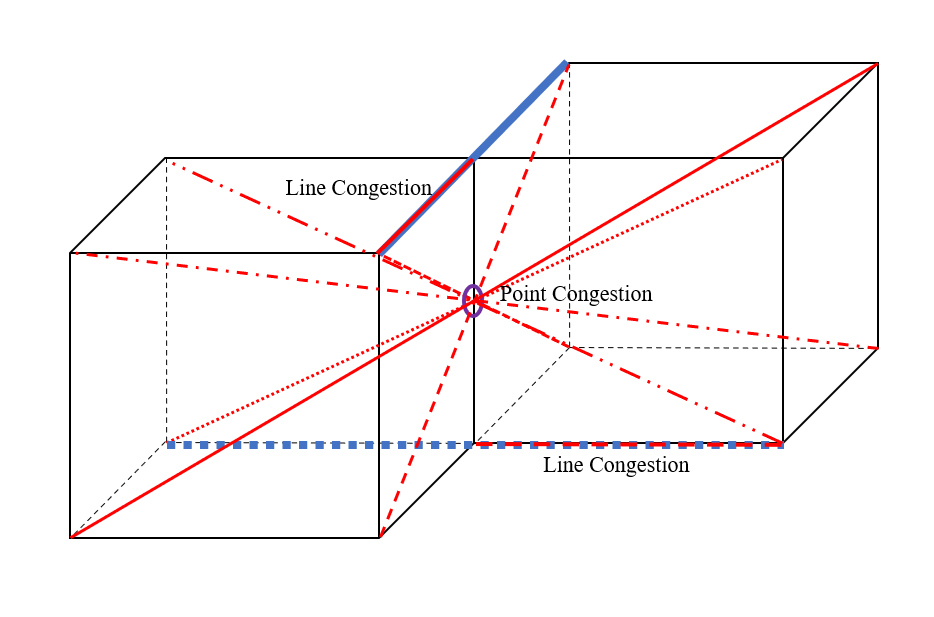}
   \caption{The more unique congestion in the edge-sharing cubic Network}
   \label{fig:7}
\end{figure}

However, in line-sharing two-cubic network and node-sharing two-cubic network, peculiar congestion points emerge. In Figure \ref{fig:7} a  \textit{new redundunt} point congestion by four planar diagonals and two spacial diagonals invades a shared line. Intuitively, we can see that a player at the intersecting node utilizing the unit shared link faces to more complicated congestion cost than the player in the plane-sharing cubic network.

\begin{figure}[H]
   \centering
       \includegraphics[scale=.65]{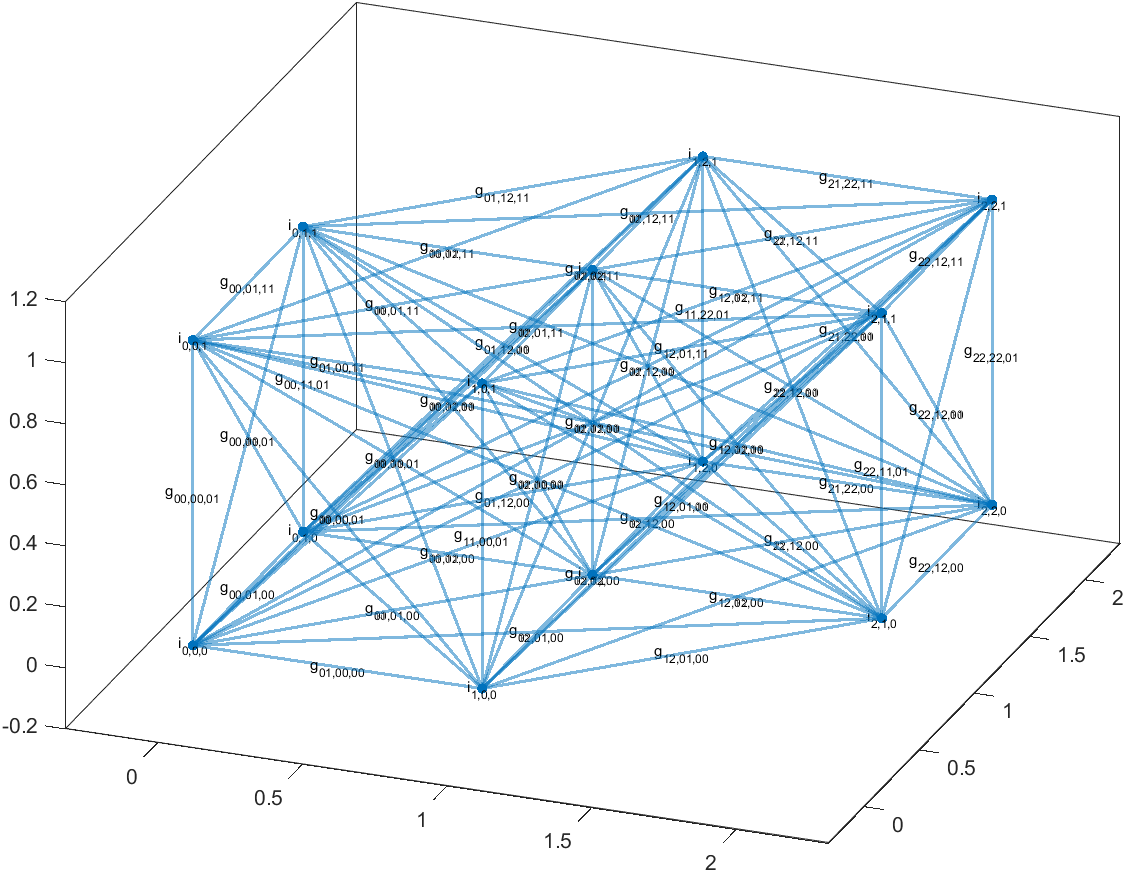}
   \caption{The complete line-sharing Network}
   \label{fig:8}
\end{figure}

Furthermore, this line-sharing two-cubic network creates new paths. We can observe in Figure \ref{fig:8} that this network can create additional new point congestion, even outside the cube. This line-sharing two-cubic network has a total of 91 paths. This out-of-two-cubes generating of point congestions can be called a new externality, not a network within the cube itself. The externality causes other congestions in neighbor cube network.

The node-sharing two-cubic network has a bizarre congestion coordinate. This congestion must be defined separately from point congestion because we defined point congestion in Definition \ref{def:2} excluding nodes. The odd congestion coordinate, which has the paradoxical property of being an intersection at a node, is a point that must be defined separately from point congestion. Furthermore, this erratic congestion coordinate starts with line congestion on all unit links, which means that the line congestion cost is default for the unit delivery paths. The bizarre coordinate needs to be defined as a new type: the full congestion.

\begin{definition}
    {\large\textbf{Full congestion}} is the congestion at a node, where two or more links intersect at a point and the delivery paths cross each other.
    \label{def:4}
\end{definition}

\begin{corollary}
    All unit links of the node fully congested are line congested.
    \label{crl:2}   
\end{corollary}

\begin{figure}[H]
   \centering
       \includegraphics[scale=.7]{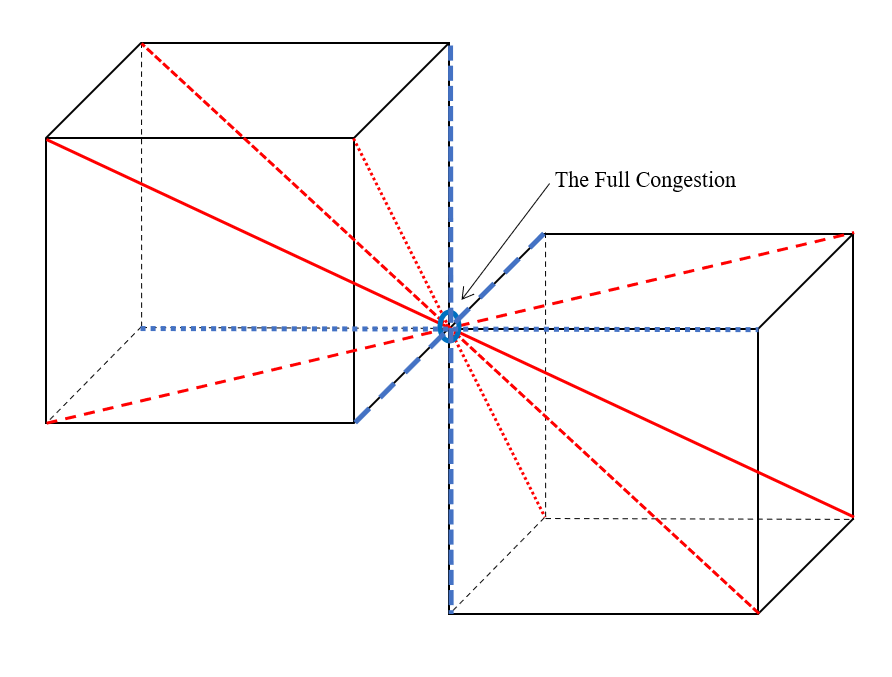}
   \caption{The node-sharing Network}
   \label{fig:9}
\end{figure}

In Figure \ref{fig:9}, we can observe how this congestion point is created.

\begin{figure}[H]
   \centering
       \includegraphics[scale=.65]{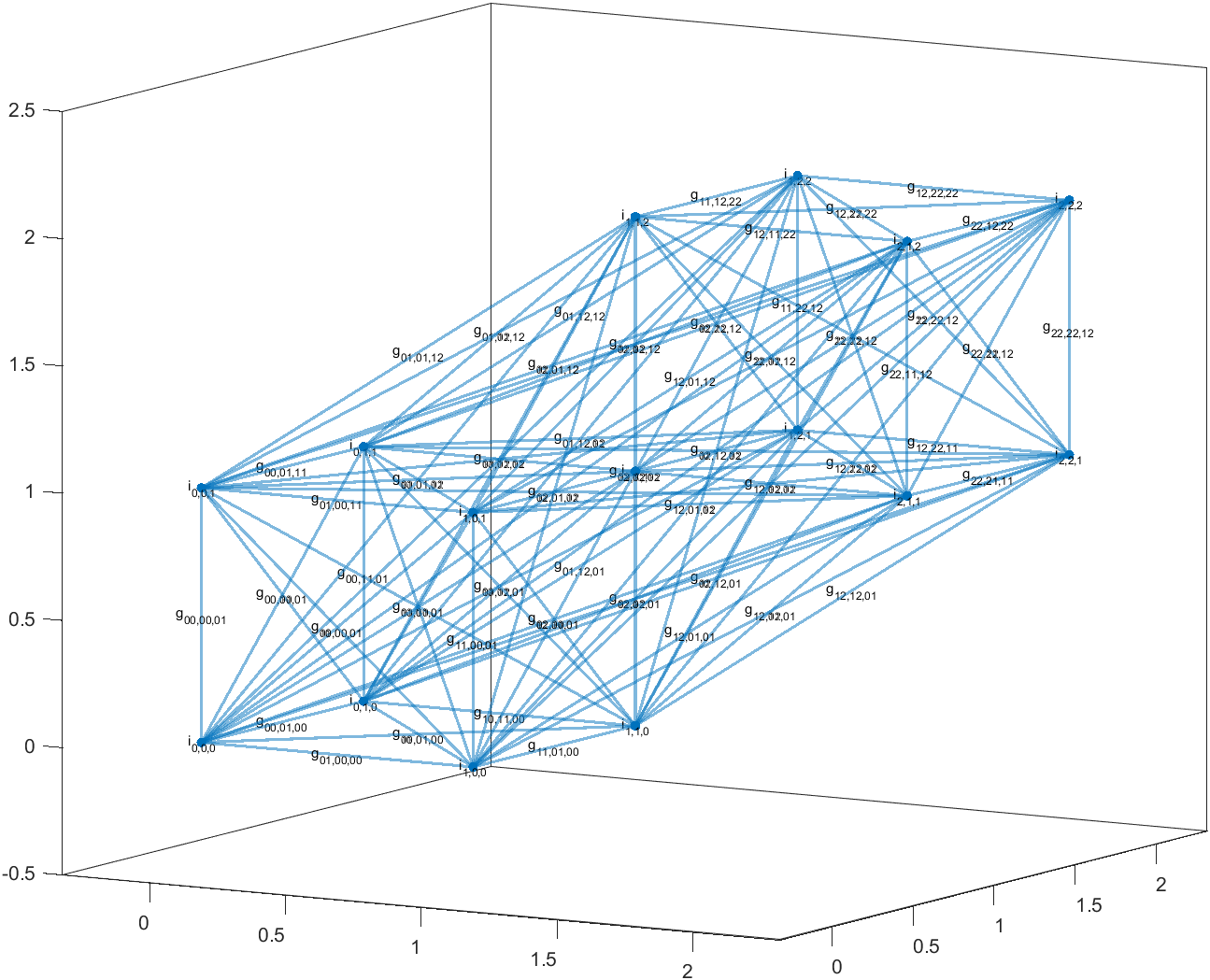}
   \caption{The complete node-sharing Network}
   \label{fig:10}
\end{figure}

The node-sharing two-cubic network has a total of 105 paths. Figure \ref{fig:10} shows all the links that this network can have. More point congestions are observed outside the cube than in the line-sharing two-cubic network. The space covered by the congested points outside the cube is also larger than in the line-sharing two-cubic network. That is, the externality of congestions are bigger.

Because the existence of full congestion in this node-sharing two-cubic network, all unit links and nodes are filled with congestion coordinates, which means that every edge and node of the cube correspond to every congestion coordinate. Above all, in the space of the single cube network, we can find any other two cubes which make a congestion coordinate. Intuitively, if you divide a cube into the infinite numbers of small cubes and draw all diagonal paths through them, you can observe that the diagonal paths overlap as line congestion and point congestion inside a centered single cube.

The proof can also be done the other way around. Consider a cube and attach other cubes that share points, lines, and faces with it, i.e., a 3x3 cube network. In a network of 27 cubes, the 8 nodes belonging to the center cube create a total of 28 links.\footnote{This can be seen in Figure \ref{fig:3}} All diagonals in the centered cube overlap with links out of the other 26 neighboring cubes, which means that all the links the centered cube creates are congestion points if it is surrounded by the adjacent cubes. Moreover, the adjacent cubes creates the externality congestion coordinates in the space within the centered cube. If this congestion generation is repeated as the number of adjacent cubes increases the interior of the center cube will be filled with congestion.

In conclusion, the theoretical examination of congestion cost in three-dimensional space leads to filling all nodes and links with congestion coordinates when considering a countless-cubic network. I will address this paradoxical phenomenon as the {\textit{congestion paradox}} in the delivery network.

\begin{theorem}
    Given a multiple cubic delivery network consisting of infinite number of cubes, {\large\textbf{congestion Paradox}} is a state when all coordinates including nodes and links are congested by different delivery paths, which guarantee the same positive payoff.
    \label{thm:2}   
\end{theorem}

\begin{corollary}
    Given a multiple cubic delivery network consisting of finite numbers of cubes, there is a coordinate which is not congested by delivery paths in some cube.
    \label{crl:3}   
\end{corollary}

\section{Further discussion and Conclusions}

The congestion paradox is a different prospect if payoff varies in proportion to the distance of paths since the congestion paradox is caused by a myriad of different paths that guarantee the same payoff. If goods are heterogeneous, the payoff can be varied as well. That is, heterogeneous goods will provide different benefits. When the heterogeneity of goods is determined by probability, individual players approach the payoff optimization problem in terms of expected utility rather than determined utility.

Cost minimization of establishing new links can also be approached as a social planner's problem. The social planner optimizes the number and type of links $g_{ll',mm',nn'}$ to maximize social welfare. The problem of congestion cost minimization subject to maximizing social welfare require a complicated mathematical analysis.

In conclusion, our economic analysis of a linear network, a planar network, and a cubic network shows that each player will randomly choose a delivery path to deliver all goods given the payoff for all delivery paths is the same. If congestion costs are introduced to the player's maximization problem in a multi-cubic delivery network, the congestion paradox arises where all coordinates become congested as long as the previous assumptions about payoffs are maintained. This congestion paradox may not occur when the costs and benefits vary, and the welfare optimization by a social planner who decides the supply of links is left as a future research issue.

\newpage

\bibliography{references.bib}

\newpage

\appendix
\appendixpage

    \section{Best response of the player regarding Kuhn-Tucker (K-T) conditions with each delivery path vector in the linear Network}
    \label{apdx:A}
The payoff function $\Pi_i = b_{ij}v_{ij}x_{ij}-c_{ij}v_{ij}x_{ij}-c_{ii}v_{ii}x_{ii}$. Referring to $\textbf{v}_{ij}$, the player chooses four different vectors in the strategy function.

i) When the player chooses $\textbf{v}_i = \begin{bmatrix}0 \\ 0 \end{bmatrix}$, then $\textbf{x}_i = \begin{bmatrix}0 \\ 0\end{bmatrix}$, which contradicts $\sum_{j}x_{ij} = 1$ for all $j$. Thus, the strategy with $\textbf{v}_i = \begin{bmatrix}0 \\ 0 \end{bmatrix}$ is excluded from the best response of the player $i$.

ii) Let $\textbf{v}_i = \begin{bmatrix}1 \\ 0 \end{bmatrix}$, which is $\textbf{x}_i = \begin{bmatrix}x_{ii}>0 \\ x_{ij}=0\end{bmatrix}$. Then, the Lagrangian function
\begin{equation*}
    \text{\Large $\mathcal{L}$} = -c_{ii}x_{ii} + \begin{bmatrix}\lambda_1 \\ \lambda_2 \\ \lambda_3 \end{bmatrix}' \begin{bmatrix}-c_{ii} \\ -c_{ij} \\ c_{ij}-b_{ij} \end{bmatrix} + \mu(1-x_{ii}-x_{ij}).
\end{equation*}
The K-T conditions with $\textbf{v}_i = \begin{bmatrix}1 \\ 0 \end{bmatrix}$ are
\begin{equation*}
\begin{aligned}
    \frac{\partial \text{\Large $\mathcal{L}$}}{\partial x_{ii}} &= -c_{ii} - \mu = 0, \;(x_{ii}>0)\\
    \frac{\partial \text{\Large $\mathcal{L}$}}{\partial x_{ij}} &= \mu < 0, \;(x_{ij}=0)\\
    {c}_{ii} &> 0, \\
    b_{ij} > c_{ij} &> 0, \\
    \sum_{j}x_{ij} &= 1, \;and \\
    \lambda_1 = \lambda_2 =\lambda_3 &= 0 
\end{aligned}
\end{equation*}
Then the payoff of the player $i$ with $\textbf{v}_i = \begin{bmatrix}1 \\ 0 \end{bmatrix}$ is $-c_{ii}x_{ii} < 0$.

iii) Let $\textbf{v}_i = \begin{bmatrix}0 \\ 1 \end{bmatrix}$, which is $\textbf{x}_i = \begin{bmatrix}x_{ii}=0 \\ x_{ij}>0\end{bmatrix}$. Then, the Lagrangian function
\begin{equation*}
    \text{\Large $\mathcal{L}$} = b_{ij}x_{ij}-c_{ij}x_{ij} + \begin{bmatrix}\lambda_1 \\ \lambda_2 \\ \lambda_3 \end{bmatrix}' \begin{bmatrix}-c_{ii} \\ -c_{ij} \\ c_{ij}-b_{ij} \end{bmatrix} + \mu(1-x_{ii}-x_{ij}).
\end{equation*}
The K-T conditions with $\textbf{v}_i = \begin{bmatrix}0 \\ 1 \end{bmatrix}$ are
\begin{equation*}
\begin{aligned}
    \frac{\partial \text{\Large $\mathcal{L}$}}{\partial x_{ii}} &= - \mu < 0, \;(x_{ii}=0)\\
    \frac{\partial \text{\Large $\mathcal{L}$}}{\partial x_{ij}} &= b_{ij}-c_{ij} - \mu = 0, \;(x_{ij}>0)\\
    {c}_{ii} &> 0, \\
    b_{ij} > c_{ij} &> 0, \\
    \sum_{j}x_{ij} &= 1, \;and \\
    \lambda_1 = \lambda_2 =\lambda_3 &= 0 
\end{aligned}
\end{equation*}
Then the payoff of the player $i$ with $\textbf{v}_i = \begin{bmatrix}1 \\ 0 \end{bmatrix}$ is $(b_{ij}-c_{ij})x_{ij} > 0$.

iv) Let $\textbf{v}_i = \begin{bmatrix}1 \\ 1 \end{bmatrix}$, which is $\textbf{x}_i = \begin{bmatrix}x_{ii}>0 \\ x_{ij}>0\end{bmatrix}$. Then, the Lagrangian function
\begin{equation*}
    \text{\Large $\mathcal{L}$} = b_{ij}x_{ij}-c_{ij}x_{ij}-c_{ii}x_{ii} + \begin{bmatrix}\lambda_1 \\ \lambda_2 \\ \lambda_3 \end{bmatrix}' \begin{bmatrix}-c_{ii} \\ -c_{ij} \\ c_{ij}-b_{ij} \end{bmatrix} + \mu(1-x_{ii}-x_{ij}).
\end{equation*}
The K-T conditions with $\textbf{v}_i = \begin{bmatrix}1 \\ 1 \end{bmatrix}$ are
\begin{equation*}
\begin{aligned}
    \frac{\partial \text{\Large $\mathcal{L}$}}{\partial x_{ii}} &= - \mu = 0, \;(x_{ii}>0)\\
    \frac{\partial \text{\Large $\mathcal{L}$}}{\partial x_{ij}} &= b_{ij}-c_{ij} - \mu = 0, \;(x_{ij}>0)\\
    {c}_{ii} &> 0, \\
    b_{ij} > c_{ij} &> 0, \\
    \sum_{j}x_{ij} &= 1, \;and \\
    \lambda_1 = \lambda_2 =\lambda_3 &= 0 
\end{aligned}
\end{equation*}
Then the payoff of the player $i$ with $\textbf{v}_i = \begin{bmatrix}1 \\ 1 \end{bmatrix}$ is $(b_{ij}-c_{ij})x_{ij}-c_{ii}x_{ii}$.

From (ii),(iii), and (iv), the largest payoff is $(b_{ij}-c_{ij})x_{ij}$ if the delivery path vector $\textbf{v}_i$ is $\begin{bmatrix}0 \\ 1 \end{bmatrix}$ with $x_{ij} =1$. Hence, the player $i$ always chooses the strategy $S_i = \begin{bmatrix}v_{ii} = 0 \\ v_{ij} = 1 \end{bmatrix} \times \begin{bmatrix}x_{ii} = 0 \\ x_{ij} = 1 \end{bmatrix}$ to maximizes his payoff in the line network.

\end{document}